\documentclass[12pt]{article}
\usepackage{latexsym,epsf}
\newcommand{\beq}[3]{\begin{equation}  \label{#1#2#3}}
\newcommand{\eeq}{ \end{equation}}
\newcommand{\ba}{\begin{array}}
\newcommand{\ea}{\end{array}}

\newcommand{\remark}[1]{}
\newcommand{\be}[3]{\begin{equation}  \label{#1#2#3}}     % non-hyper
\newcommand{\ee}{ \end{equation}}
\newcommand{\bea}{\begin{eqnarray}}
\newcommand{\eea}{\end{eqnarray}}

\newcommand{\ft}[2]{{\textstyle\frac{#1}{#2}}}

\def\beq{\begin{equation}}
\def\eeq{\end{equation}}
\def\beqa{\begin{eqnarray}}
\def\eeqa{\end{eqnarray}}

\begin{document}

\begin{titlepage}
\rightline{HU-EP-02/25}
\rightline{hep-th/0207024}

\vspace{15truemm}

\centerline{\bf \LARGE
The holographic RG flow in a
}
\vskip0.3cm
\centerline{\bf \LARGE
field theory on a curved background
}

\bigskip

\vspace{2truecm}

\centerline{\bf Gabriel Lopes Cardoso
{\rm and} Dieter L\"ust}

\vspace{1truecm}

\centerline{\em  Institut f\"ur Physik, Humboldt University}
\centerline{\em Invalidenstra\ss{}e 110, 10115 Berlin, Germany}
\vspace{1truecm}
\centerline{\tt email:gcardoso,luest@physik.hu-berlin.de}

\vspace{2truecm}

%%%%%%%%%%%%%%%%%%%%%%%%%%%%%%%%%%%%%%%%%%%%%%%%%%%%%%%%

\begin{abstract}

\noindent
As shown by Freedman, Gubser, Pilch and Warner, the RG flow
in ${\cal N}=4$ super-Yang-Mills theory broken to an ${\cal N}=1$
theory by the addition of a mass term can be described in terms of
a supersymmetric domain wall solution in five-dimensional ${\cal N}=8$
gauged supergravity.  The FGPW flow is an example of a holographic RG flow
in a field theory on a flat background.  Here we put the
field theory studied by  Freedman, Gubser, Pilch and Warner on a curved
$AdS_4$ background, and we construct the supersymmetric domain wall
solution which describes the RG flow in this field theory.
This solution is a curved (non Ricci flat)
domain wall solution.
This example demonstrates that holographic RG flows in
supersymmetric field theories
on a curved $AdS_4$ background can be described in terms of curved
supersymmetric domain wall solutions.

\end{abstract}

\end{titlepage}

%%%%%%%%%%%%%%%%%%%%%%%%%%%%%%%%%%%%%%%%%%%%%%%%%%%%%%%%%%%%%%%%

\newpage
\section{Introduction}

The AdS/CFT
correspondence \cite{Maldacena:1997re,Gubser:1998bc,Witten:1998qj},
together with
the domain wall/QFT correspondence \cite{BSK}, states that
renormalization
group (RG) flows in field theories may be described in terms
of domain wall solutions in dual gauged supergravity theories.
An example of such a holographic RG flow has been constructed
by Freedman, Gubser, Pilch and Warner \cite{FGPW}.  This is a flow
in ${\cal N}=4$ super-Yang-Mills theory broken to an ${\cal N}=1$
theory by the addition of a mass term for one of the three adjoint
chiral superfields \cite{KLM}.  Its dual description is in terms of a
supersymmetric domain wall solution in five-dimensional ${\cal N}=8$
gauged supergravity.
This  domain wall is supported by
two non-constant scalar
fields, which are in one-to-one correspondence with the fermionic and
bosonic mass terms in the dual field theory.
The ${\cal N}=2$ version of this solution has been
given in \cite{CDKV}.

Holographic RG flows have, so far, only been studied in field theories
on a flat background.  Here we put the field theory studied by
Freedman, Gubser, Pilch and Warner on a curved $AdS_4$ background and
we show that the RG flow in this field theory has a dual description
in terms of a curved (non Ricci flat) supersymmetric domain wall solution in
five-dimensional gauged supergravity.  We construct this curved domain
wall solution in the context of ${\cal N}=2$ gauged supergravity in
five dimensions.  This demonstrates that
holographic RG flows in
supersymmetric field theories
on a curved $AdS_4$ background can be described in terms of curved
supersymmetric domain wall solutions.

In \cite{CDL2} we gave a general recipe for
constructing curved supersymmetric domain wall solutions
in the context of five-dimensional ${\cal N}=2$ gauged supergravity
with vector and hypermultiplets.  We also explicitly constructed an example
of a curved domain wall solution in a gauged supergravity model
with one hypermultiplet.
Related work appeared in \cite{CDL1,CSabra,BC}.
In \cite{CDL2} we
also discussed the dual description
of these curved BPS domain wall solutions in terms of RG flows.  We proposed
that 1) curved BPS domain wall solutions may provide a dual description
of RG flows in field theories on a curved background with $AdS_4$
curvature and 2) that the curvature on the domain wall
may act as an infrared regulator in the dual field theory.  Here
we provide evidence for 1).

The curved BPS flow equations describing the curved version of the FGPW flow
are complicated and difficult to solve.  We therefore restrict ourselves to
constructing the curved domain wall solution in the vicinity of the
UV fixed point.  We show that the curved BPS domain wall solution is
supported by more scalar fields than its flat counterpart,
and that the scalar fields (which involve both vector and hyper scalar
fields)
supporting the curved domain wall solution
are in one-to-one
correspondence with mass operators in the dual field theory, some
of which are induced by putting the field theory on a curved background.

\section{A curved version of the FGPW
domain wall solution}

\setcounter{equation}{0}

The flat domain wall solution constructed by Freedman, Gubser, Pilch and
Warner is a solution to five-dimensional ${\cal N}=8$ gauged supergravity,
and it interpolates between two $AdS$ vacua.  These two vacua correspond
to two RG fixed points in the dual field theory, one in the UV regime
and the other one in the IR regime.

The FGPW solution can also be described in the context of ${\cal N}=2$
gauged supergravity in five dimensions \cite{CDKV}.  In the following,
we will construct a curved version of this flat
domain wall solution.  The five-dimensional ${\cal N}=2$ gauged
supergravity theories that we consider are in the class constructed
in \cite{AnnaGianguido},
describing the general coupling of vector and hypermultiplets
to supergravity.

The flat domain wall solution constructed in \cite{CDKV} arises in a
five-dimensional
${\cal N}=2$ gauged supergravity theory with scalar manifold
\bea
{\cal M} = O(1,1) \times \frac{SU(2,1)}{SU(2) \times U(1)} \;.
\eea
The $O(1,1)$ factor denotes the vector scalar manifold, and it is
parametrised by one vector scalar $\rho$.  The metric of this
very special manifold
is given by $g_{\rho \rho}= 12 \rho^{-2}$.
The factor
$\frac{SU(2,1)}{SU(2) \times U(1)}$, on the other hand, denotes a
quaternionic K\"ahler space which is parametrised by four scalar fields
belonging to a hypermultiplet.  These four scalar fields are denoted
by $q^X = (V, \sigma, \theta, \tau)$ and the associated line element
reads $ds^2 = \ft12 V^{-2} dV^2 + \ft12 V^{-2} (d \sigma - 2 \tau d \theta
+ 2 \theta d \tau)^2 + 2 V^{-1} ( d \theta^2 + d \tau^2)$.

The flat domain wall solution of FGPW is then obtained by performing a
gauging of two compact isometries of the quaternionic K\"ahler space
described above.
This gauging results in the following
triplet of Killing prepotentials \cite{CDKV} (we refer to \cite{CDKV} for
the details)
\bea
P^s &=& \sqrt{\frac{3}{2}} \left(\frac{2 \theta + V \theta \rho^6 -
\theta^3 \rho^6 - \sigma \tau \rho^6  - \theta \tau^2 \rho^6}{
\sqrt{V}  \rho^2},
\frac{\sigma \theta  \rho^6 + \tau
(2 + V \rho^6 - \theta^2 \rho^6 - \tau^2 \rho^6 )}{
\sqrt{V} \rho^2}, \right.\nonumber\\
&&\left.
\frac{4(- V +
\theta^2 + \tau^2) - \rho^6 (1 + \sigma^2 +
                    V^2 + \theta^4 +
                    2 \theta^2 \tau^2 + \tau^4 -
                    6 V (\theta^2 + \tau^2))}{4 V \rho^2}
\right) .
\eea
This triplet of Killing prepotentials may be decomposed
into its norm $W$ and into $SU(2)$ phases $Q^s$ according to
$P^s = \sqrt{\ft32} W Q^s$ with $Q^s Q^s =1 $.
The associated $W^2$ is then given by
\bea
\label{w2}
W^2 &=& \frac{1}{16 V^2 \rho^4} \left(
16 V [\sigma \theta
\rho^6 + \tau (2 + V \rho^6 - \theta^2  \rho^6 -
\rho^6 \tau^2)]^2  \right. \\
  &&    + 16 V [\theta^3 \rho^6 + \sigma \rho^6 \tau +
\theta (-2 - V \rho^6 + \rho^6 \tau^2)]^2
 + [
-4 (-V + \theta^2 + \tau^2) \nonumber\\
 &&\left. + \rho^6 (1 + \sigma^2
+ V^2 + \theta^4 + 2 \theta^2 \tau^2 + \tau^4 - 6 V
(\theta^2 + \tau^2))]^2 \right) \;,\nonumber
\eea
and
the triplet $Q^s$ is given by $Q^s = \sqrt{\ft23} P^s/W$.

The flat domain wall solution constructed in \cite{CDKV} is supported
by the vector scalar $\rho$ and by the hyper scalar field $\tau$.
More precisely, the hyper scalars supporting this flat solution
satisfy \cite{CDKV}
\bea
\sigma = 0
\label{sig}
\eea
as well as
\bea
\xi^2 = \theta^2 + \tau^2 \;\;\;,\;\;\; \xi = \sqrt{1 -V} \;\;\;,\;\;\;
0< V \leq 1\;.
\label{xif}
\eea
We may set
\bea
\theta = 0
\label{thef}
\eea
without loss of generality.
The UV fixed point is at $\rho = V =1$, while the IR fixed point is
at $\rho = 2^{1/6}, V= \ft34$.
In the dual field theory
(and using the ${\cal N}=1$ component notation of \cite{WB}),
a non-constant
vector scalar $\rho$ corresponds to the addition of
a mass term for one of the adjoint scalar fields, $\rho\leftrightarrow
m^2A\bar A$,
whereas a non-constant hyper scalar field $\tau$ corresponds to the
associated (${\cal N}=1$ supersymmetric)
fermionic mass term, $\tau\leftrightarrow
m\chi\chi$.

We now proceed to construct a curved version of this
flat domain wall solution.  We write the five-dimensional line element as
\bea
ds^2 = {\rm e}^{2 U(r)} {g}_{mn} \, dx^m dx^n + dr^2 \,
\eea
where the metric ${g}_{mn}$ is taken to be a four-dimensional
constant curvature metric satisfying ${R}_{mn} = 12 \Lambda^2
g_{mn}$, and where $\Lambda$ denotes a real constant.  This
corresponds to a four-dimensional anti-de Sitter spacetime.

There is a characteristic quantity which enters in the construction of
curved supersymmetric domain wall solutions, and which is given by
\cite{DeWolfe:2000cp,CDL2}
\bea
\gamma = \sqrt{1 - 4 \Lambda^2 ({\rm e}^U g W)^{-2}} \;.
\label{gamma}
\eea
A flat domain wall satisfies $\Lambda =0$.  A curved version will receive
corrections to all orders in $\Lambda$, as suggested by power expanding
$\gamma$ in $\Lambda$.  Here we will restrict ourselves to constructing
a curved domain wall solution to order $\Lambda^2$.

The curved domain wall solution will be supported by some of the scalar fields
$\rho,V, \sigma, \theta$ and $\tau$ .
As in the flat case, we again set
\bea
\sigma = 0 \;\;\;,\;\;\; \theta =0 \;.
\label{st0}
\eea
We will check that the restriction (\ref{st0})
is consistent with both the equations of motion
and the curved BPS flow equations for the scalar fields.
We also write
\bea
\tau = \sqrt{1 - V + f} \;,
\label{simpl}
\eea
for convenience.
%We note that $(1 - V  + f) \geq 0$.
The curved domain wall will thus be supported
by the three scalar fields $\rho (r), \tau (r)$ and $f(r)$.

According to the general recipe given in \cite{CDL2} for constructing
curved BPS domain wall solutions, we first have to specify a triplet
$M^s$ satisfying $M^s M^s =1, M^s Q^s =0$.
In addition, $M^s$ has to be consistent with
\cite{CDL2}
\bea
B M^s = \mp \gamma \frac{\partial_{\rho} Q^s}{W^{-1} \partial_{\rho} W} \;,
\label{normb}
\eea
where $B = \sqrt{1 - \gamma^2}
= 2 \Lambda ({\rm e}^{U} gW)^{-1}$.
We take
\bea
M^s = \frac{1}{[(Q^2)^2 + (Q^3)^2]} \, (0, - Q^3, Q^2 ) \;.
\label{m}
\eea
For domain walls satisfying (\ref{st0}) we find that $\partial_{\rho}
Q^s \propto M^s$, so that (\ref{m}) and (\ref{normb}) are indeed
consistent with one another when subjected to (\ref{st0}).

{From} (\ref{normb}) we obtain (we choose the upper sign in the following)
\bea
B \gamma^{-1} = -
\frac{M^s \partial_{\rho} Q^s}{W^{-1} \partial_{\rho} W}
= 12 \alpha^{-1}
f (2+f) \sqrt{1-V+f} \sqrt{V} \rho^6 , \label{bf}
\eea
where
\bea
\alpha &=&
8 (1+f)^2 + 2 (2+f(2+f)) (1+f-2V)\rho^6 - (2+f(2+f))^2 \rho^{12} \nonumber\\
&+& 16 (1+f) V \rho^{12} - 16 V^2 \rho^{12} \;.
\label{alpha}
\eea
On the other hand, since
\bea
B \gamma^{-1}=  2 \Lambda (\gamma {\rm e}^{U} gW)^{-1} \;,
\label{bg}
\eea
we see from (\ref{bf})
that whenever the wall is curved ($\Lambda \neq 0$) we have $f \neq 0$ and
$\tau \neq 0$
(away from the fixed points).

As mentioned above, we will restrict ourselves to
constructing the curved domain wall solution to
order $\Lambda^2$.
To this order we
obtain $2 \Lambda ({\rm e}^U g W)^{-1}$ for the lhs of (\ref{bf}).

Since $f$ if of order $\Lambda$ and higher, we
expand the rhs of (\ref{bf}) in powers of $f$.  To linear
order in $f$ we then obtain
\bea
f = - \frac{2}{3g} \Lambda \, {\rm e}^{-U_0} \, \sqrt{\frac{V_0}{1-V_0}}
\, \rho_0^{-4} \,
[-1 + (- 1 + 2 V_0)\, \rho_0^6 ] \;,
\label{f1}
\eea
where we used (\ref{st0}).
Here the subscript $0$ refers to the flat solution.
This thus determines
$f$ to lowest order in the cosmological constant $\Lambda$
on the wall\footnote{
%{\bf A word of caution:}
In deriving (\ref{f1}) we set $\gamma \approx 1$.  This is consistent
as long as ${\rm e}^{-2U}$ stays finite.  In the infrared, however,
${\rm e}^{-2U}$ diverges in which case the approximation $\gamma \approx 1$
breakes down.  Thus,
(\ref{f1}) cannot be trusted in the infrared.}.
To quadratic order in $f$, on the other hand,
we obtain from (\ref{bf})
\bea
\Lambda {\rm e}^{-U} = &-&
\frac{3g}{2} \, \frac{\sqrt{1-V}}{\sqrt{V}}\, \rho^4 \, \frac{f}{
[-1 + (-1+2V) \,\rho^6]} \nonumber\\
&+& \frac{3g}{4} \, \frac{\sqrt{V}}{\sqrt{1-V}} \, \rho^4 \,
\frac{[1 + (-3+2V)\, \rho^6] \,f^2}{[-1 + (-1 + 2V) \,\rho^6]^2} \;,
\label{lff}
\eea
where we again used (\ref{st0}).  Using (\ref{f1}) and
(\ref{lff}), we find that, to quadratic order in $\Lambda$, $f$ is given by
\bea
\label{f2}
f = &-& \frac{2}{3g} \Lambda \, {\rm e}^{-U} \, \sqrt{\frac{V}{1-V}}
\, \rho^{-4} \,
[-1 + (-1+2 V ) \,\rho^6]   \\
&+& \frac{2}{9g^2}
\Lambda^2 \, {\rm e}^{- 2 U_0} \, \frac{V_0^2}{(1-V_0)^2} \, \rho_0^{-8} \,
[1 + (-3+2V_0) \, \rho_0^6] [-1 + (-1+2 V_0 ) \,\rho_0^6] \nonumber\\
&+& {\cal O} (\Lambda^3)
\;.
\nonumber
\eea
Here,
$U, V$ and $\rho$ denote the order $\Lambda$-corrected
expressions, whereas $U_0, V_0$ and $\rho_0$ denote the expressions
appearing in
the flat domain wall solution.  The expressions for $U, V$ and $\rho$ will
be determined by solving the associated curved BPS equations for these
fields.
The expression (\ref{f2}) for $f$, on the other hand,
should solve the curved BPS
equation for this field to order $\Lambda^2$.  This will indeed turn out to
be the case.

Let us now check that the curved domain wall specified by (\ref{st0}),
(\ref{m}) and (\ref{f2}) satisfy the scalar equations of motion given in
\cite{CDL2}.  The equation of motion for the vector scalar $\rho$ reads
(we again take the upper sign)
\bea
 \Big( \rho^{ \prime}\,  \partial_{\rho}
W  \, \partial_{\rho} \Gamma - \gamma^{\prime}
\partial_{\rho} W \Big) + \gamma^2 q^{X \prime} \Big( {\Sigma_X}^Y
\partial_{\rho} \partial_Y W + \gamma^{-1} \partial_{\rho} \partial_X W
\Big) =0 \; ,
\label{mov}
\eea
where the prime denotes the derivative with respect to $r$, and where
\bea
\Gamma^{-2} &=& 1 + W^2 \,
\frac{(\partial_{\rho} Q^s )\, (\partial_{\rho} Q^s)}{(\partial_{\rho} W) \,
(\partial_{\rho} W)} \;, \nonumber\\
{\Sigma_X}^Y &=& -\gamma \,
{\delta_X}^Y
+ 4 \Lambda ({\rm e}^U g W)^{-1} \,
\varepsilon^{rst} M^r Q^s {R^t_X}^Y \;.
\eea
Working to order $\Lambda^2$, we find that (\ref{mov}) contains
terms proportional to $f^2$ and to $\Lambda^2$.
To arrive at this result,
we have made use of the curved BPS flow equations which we will discuss
below (c.f. (\ref{floweqs})).  We find that the terms proportional to
$f^2$ and to $\Lambda^2$ precisely
cancel out by virtue of (\ref{f1})!  The equations of motion for the
hyper scalar fields are given by
\bea
\label{moh}
&& \gamma^{-2}  \rho^{\prime}\,  \partial_{\rho} W  \, \partial_Z \Gamma
- 4 g \gamma W {\Sigma_Z}^Y \partial_Y W - 4 g W \partial_Z W \\
&& + q^{X \prime} (\partial_Z {\Sigma_X}^Y ) \partial_Y W
%- \partial_X \Sigma_Z\,^Y \Big) \partial_Y W
- {\Sigma^\prime_Z}^Y \partial_Y W \nonumber\\
&& - q^{X \prime} \Big( {\Sigma_Z}^Y \partial_X \partial_Y W
- {\Sigma_X}^Y \partial_Z \partial_Y W \Big)
- \rho^{\prime} \Big(  \gamma^{-1} \partial_Z \partial_{\rho} W +
{\Sigma_Z}^Y \partial_Y \partial_{\rho} W \Big) = 0 \;. \nonumber
\eea
Making again use of the curved BPS flow equations we find that
to order $\Lambda^2$ (\ref{moh}) contains terms proportional
to $f, \Lambda, f^2$ and to $\Lambda^2$.  All these terms precisely
cancel out by virtue of the relations (\ref{f1}) and (\ref{f2})!

Next, let us turn to curved BPS flow equations for the warp factor
and for the scalar fields.
These are given by \cite{CDL2}
\bea
U^\prime &=&  \gamma g W \;,
 \nonumber\\
\rho^{\prime} &=&  - \frac{g}{4} \,
\gamma^{-1} \, \rho^2 \, \partial_{\rho} W\;,
\nonumber\\
q^{X \prime} &=& 3 g\, g^{XY} {\Sigma_Y}^Z \partial_Z W \;,
\label{floweqs}
\eea
where, again, we chose the upper sign.
We find that (\ref{st0}) is a
solution to (\ref{floweqs}) to all orders in $\Lambda$.

Let us first solve (\ref{floweqs}) to linear order in
the cosmological constant $\Lambda$.
To this order,
the BPS flow equation for $f$ is given by
\bea
f^{\prime} &=& 3g \left(\sqrt{1-\gamma^2}\,  \sqrt{\frac{1-V}{V}}\,
 \frac{(\rho^6-2)}
{\rho^2} - f \, \rho^4 \, \frac{(4V + \rho^6 -2)}{2 + \rho^6(2V-1)}
\right)  \nonumber\\
&=&
2 \Lambda \, {\rm e}^{-U_0} \, \sqrt{\frac{V_0}{1-V_0}} \, (-5 + 4 V_0 +
\rho_0^6)  \;,
\label{fprime1}
\eea
where we used (\ref{f1}) to rewrite the rhs.

On the the other hand, we can also compute $f^{\prime}$ directly
from (\ref{f1}).
Using the flat BPS equations
\bea
U_0^{\prime} &=& g\, \frac{(2 + \rho_0^6 (2V_0-1))}{2 V_0 \rho_0^2} \;,
\nonumber\\
V_0^{\prime} &=& 3 g \,\frac{(V_0-1)(\rho_0^6 -2)}{\rho_0^2} \;, \nonumber\\
\rho_0^{\prime} &=& g \, \frac{[1+\rho_0^6(1 -2V_0)]}{
2 V_0 \rho_0} \;,
\label{bpsflat}
\eea
we find that the expression for $f^{\prime}$ computed from (\ref{f1})
is identical to (\ref{fprime1})!  That is, to lowest order in the cosmological
constant, the BPS flow equation for $f$ is solved by (\ref{f1}).

Next, let us solve the curved
BPS flow equations for the fields $U, V$ and $\rho$
to linear order in $\Lambda$.
These
are
given by
\bea
U^{\prime} &=& g \left(  \frac{2 + \rho^6 (2 V-1)}{2 V \rho^2}
+ f \frac{(2 -\rho^6) }{2 V \rho^2}
\right) \;,\nonumber\\
V^{\prime} &=& \frac{3 g}{\rho^2} \,\Big(
(V-1 + \sqrt{1 -\gamma^2} \sqrt{1-V} \sqrt{V}
)(\rho^6 -2) \nonumber\\
&&- f \frac{(-4 + \rho^6(4-\rho^6 + V (-6 + 4 V + 3 \rho^6)))}{2 + (-1 + 2V)
\rho^6}
\Big)\;, \nonumber\\
\rho^{\prime} &=& g \left( \frac{2 + \rho^6(2V-1)(-1+\rho^6(1-2V))}{
2 V \rho (2 + \rho^6 (-1 + 2V))} + f \frac{(1 + \rho^6)}{2 V \rho}
\right) \;,
\label{bpsdef}
\eea
where $f$ is given by (\ref{f1}).
These flow equations are quite complicated.  In the following we restrict
ourselves to solving them
in the vicinity of the UV fixed point.
We set
\bea
g = \frac{2}{3}
\eea
to make contact with the results of \cite{FGPW}.
Near the UV fixed point, $gW =1$ and hence
$U = r$ as well as $\sqrt{1- \gamma^2} = 2 \Lambda {\rm e}^{-r}$.
We will need the explicit form of the flat domain wall
solution in the following, which is given by \cite{FGPW}
\bea
V_0 &=& 1- \delta V_0 \;\;\;,\;\;\;
\delta V_0 = a_0^2 \,{\rm e}^{-2r} \;, \nonumber\\
\rho_0 &=& 1 + \delta \rho_0 \;\;\;,\;\;\; \delta \rho_0
= \frac{2}{3} \,a_0^2\,
{\rm e}^{-2r}\, r + \frac{a_1}{\sqrt{6}} \,{\rm e}^{-2r} \;.
\label{flats}
\eea
By inserting (\ref{flats}) into (\ref{f1}), we find that
\bea
f = 2 \Lambda \, a_0 \, {\rm e}^{-2r} \Big(1- \sqrt{\frac{3 }{2}} \,
\frac{a_1}{a_0^2}
- 2r \Big)
\label{fc}
\eea
near the UV fixed point.

Near the UV fixed point, we infer from (\ref{bpsdef}) that
the curved flow equations for $V$ and $\rho$ are, to
lowest order in $\Lambda$, given by
\bea
(\delta V)^{\prime} &=& - 2 \delta V + 4
 \Lambda \, {\rm e}^{- r} \, \sqrt{\delta V_0} \;,
\nonumber\\
(\delta \rho)^{\prime} &=& \frac{2}{3} \Big( \delta V + f \Big)
- 2 \delta \rho \; ,
\label{bpslin}
\eea
where, as before,
\bea
V &=& 1 - \delta V \;, \nonumber\\
\rho &=& 1 + \delta \rho \;.
\eea
The equations (\ref{bpslin}) are solved by
\bea
\label{sol1}
\delta V &=&  \left (a_0^2  + \Delta_V (\Lambda)
+ 4 \Lambda a_0 \, r
\right)
\, {\rm e}^{-2r} \;, \\
\delta \rho &=& \Big( \frac{2}{3} a_0^2 \,r + \frac{a_1}{\sqrt{6}} \Big) \,
{\rm e}^{-2r} + \frac{4}{3} \Lambda \, a_0 \, \Big(1 -\sqrt{\frac{3}{2}} \,
\frac{a_1}{a_0^2} \Big) \,r\, {\rm e}^{-2r}  + \Big(
\frac{2}{3} \, \Delta_V (\Lambda) \, r + \Delta_{\rho} ( \Lambda) \Big)
{\rm e}^{-2r} \;,
\nonumber
\eea
where $\Delta_V (\Lambda)$
and $\Delta_{\rho} ( \Lambda) $
denote integration constants satisfying
$\Delta_V =\Delta_{\rho}=0$ for $\Lambda =0$, which will be determined below.

Let us now solve the BPS flow equations (\ref{floweqs}) to
order $\Lambda^2$.
To order $\Lambda^2$ and to second order in $f$,
the BPS flow equation for $f$ is given by
\bea
f^{\prime} &=& 2 \left(\sqrt{1-\gamma^2}\,  \sqrt{\frac{1-V}{V}}\,
 \frac{(\rho^6-2)}
{\rho^2} - f \, \rho^4 \, \frac{(4V + \rho^6 -2)}{2 + \rho^6(2V-1)}
\right.
\nonumber\\
&& \left.- \sqrt{1-\gamma^2} \, f\, \frac{
(-3 + 2 V) (-2 + \rho^6) }{2 \sqrt{(1 - V)V} \rho^2}
\right.
\nonumber\\
&& \left.
- f^2 \, \frac{\rho_0^4
[-4 - 8 V_0 +
      4(1 + 2 (-1 + V_0) V_0)\rho_0^6 + (-1 +
            6 V_0) \rho_0^{12}]
}{2 (2 + (-1 + 2 V_0) \rho_0^6)^2}
\right)  \;.
\label{fp2}
\eea
Here
$V$ and $\rho$ denote the order $\Lambda$-corrected
expressions satisfying (\ref{bpsdef}),
whereas $V_0$ and $\rho_0$ denote the expressions
satisfying the flat BPS equations (\ref{bpsflat}).
The expression for $f$, on the other hand,
is given by (\ref{f2}).

$f^\prime$ can also be computed
directly from (\ref{f2}).  By using the
BPS equations (\ref{bpsflat})
and (\ref{bpsdef}) we find that the expression for
$f^\prime$ computed from (\ref{f2}) is identical to (\ref{fp2})!
Thus, to quadratic order in the cosmological constant, the BPS equation
(\ref{fp2})
for $f$ is solved by (\ref{f2}).

To second order in $f$ and in $\Lambda$, the BPS flow equation for
$U,V$ and $\rho$ are
given by
\bea
\label{bpssec}
U^{\prime} &=&
\gamma \, \frac{2 + \rho^6 (2 V-1)}{3 V \rho^2}
+ f \frac{(2 -\rho^6) }{3 V \rho^2}
+ f^2 \frac{\rho^4 (-2 + 4 V + \rho^6) }{6 V (2 + \rho^6 (2 V-1))}
 \;, \nonumber\\
V^{\prime} &=& \frac{2}{\rho^2} \,
[\gamma \, (V-1) + \sqrt{1 -\gamma^2} \sqrt{1-V} \sqrt{V}
](\rho^6 -2) \nonumber\\
&&- 2
f \,\frac{[-4 + \rho^6(4-\rho^6 + V (-6 + 4 V + 3 \rho^6))]}{\rho^2
(2 + (-1 + 2V)
\rho^6)}
\nonumber\\
&&- f \sqrt{1 - \gamma^2}  \frac{ \sqrt{V}
[4 - 8 (-1  + V)^2 \rho^6 + (3 - 4 V) \rho^{12}]}{
 \sqrt{1 - V} \rho^2 (2 + (-1 + 2 V) \rho^6)} \\
&& - f^2
\; \frac{\rho^4 \, [4 - 24 V^2 -
      4 (1 + 2 (-1 + V) V) \rho^6 + (1 - 4 V +
            6 V^2) \rho^{12}]}{(2 + (-1 + 2 V) \rho^6)^2 } \;,\nonumber\\
\rho^{\prime} &=&
\frac{1 + \rho^6-2V \rho^6 }{
3 \gamma V \rho } + f \frac{(1 + \rho^6)}{3 V \rho}
%\right.
%\nonumber\\
%&& \left.
+ f^2 \frac{\rho^5 [4 V^2 \rho^6 + (-2 + \rho^6)^2 -
        2 V (4 + 2\rho^6 + \rho^{12})]}{6 V
(2 + (-1 + 2 V) \rho^6)^2}
 \;. \nonumber
\eea
As before, we restrict ourselves to solving these BPS equations in
the vicinity of the UV fixed point, where
\bea
U = r + \Lambda^2 {\rm e}^{-2r} \;\;\;,\;\;\;
V = 1 - \delta V \;\;\;,\;\;\; \rho = 1 + \delta \rho \;\;\;,\;\;
\sqrt{1 - \gamma^2} = 2 \Lambda {\rm e}^{-r} \;.
\eea
Near the UV fixed point $\delta V, \delta \rho$ and $f$
are, to leading order, proportional to ${\rm e}^{-2r}$.
To order  ${\rm e}^{-2r}$, the BPS equations (\ref{bpssec}) reduce to
\bea
(\delta V)^{\prime} &=& - 2 \delta V + 4 \Lambda {\rm e}^{-r} \sqrt{\delta V}
+ 2 \Lambda {\rm e}^{-r} \frac{f}{\sqrt{\delta V}} \;, \nonumber\\
(\delta \rho)^{\prime} &=& \frac{2}{3} \, ( \delta V + f ) - 2 \delta \rho \;.
\label{vrfsec}
\eea
{From} (\ref{f2}), on the other hand, we infer that, to order  ${\rm e}^{-2r}$,
$f$ is given by
\bea
f = \Lambda {\rm e}^{-r} \left( 2 \sqrt{\delta V} - 6
\frac{\delta \rho}{\sqrt{\delta V}} \right)
+ \Lambda^2 {\rm e}^{-2r}
\left(2 - 18 \left(\frac{\delta \rho_0}{\delta V_0} \right)^2 \right) \;,
\label{f2sec}
\eea
where $\delta V_0$ and $\delta \rho_0$ are given by (\ref{flats}).
Inserting (\ref{f2sec}) into (\ref{vrfsec}) yields
\bea
\label{vrsec}
(\delta V)^{\prime} &=& - 2 \delta V + 4 \Lambda {\rm e}^{-r} \sqrt{\delta V}
+ 4 \Lambda^2 {\rm e}^{-2r}
\left(1 - 3 \, \frac{\delta \rho}{\delta V} \right) + {\cal O} (\Lambda^3)
\;, \\
(\delta \rho)^{\prime} &=& \frac{2}{3} \, \delta V - 2 \delta \rho
+ \frac{4}{3} \Lambda {\rm e}^{-r} \left( \sqrt{\delta V} - 3
\frac{\delta \rho}{\sqrt{\delta V}} \right)
+ \frac{4}{3} \, \Lambda^2 {\rm e}^{-2r}
\left(1 - 9 \left(\frac{\delta \rho_0}{\delta V_0} \right)^2 \right)
\;.
\nonumber
\eea
The equations (\ref{vrsec}) are solved by
\bea
\delta V &=&   \left(a_0^2 + \Sigma^{(1)} + \Sigma^{(2)}
- 2 \Lambda \, a_0 \Big(1- \sqrt{\frac{3 }{2}} \,
\frac{a_1}{a_0^2} \Big)
- \frac{\Lambda \Sigma^{(1)}}{a_0}
\Big(1+ \sqrt{\frac{3 }{2}} \,
\frac{a_1}{a_0^2} \Big)  + 6 \frac{\Lambda
\Delta_{\rho}}{a_0} \right.
\nonumber\\
&& \left.+ 4 \Lambda a_0 \Big(1 +  \frac{\Sigma^{(1)}}{2 a_0^2}\Big) \,
r
\right) \, {\rm e}^{-2r} \;, \nonumber\\
\delta \rho &=& \left( \frac{2}{3} \Big(a_0^2 + \Sigma^{(1)}
+ \Sigma^{(2)}
\Big) \,r + \frac{a_1}{\sqrt{6}}
+ \Delta_{\rho} \right) \,
{\rm e}^{-2r} \;,
\label{sol2}
\eea
where $\Sigma^{(1)}$ and $\Sigma^{(2)}$ denote two integration constants
of order $\Lambda$ and $\Lambda^2$, respectively, and where $\Delta_{\rho}$
denotes a third integration constant of order $\Lambda$ and higher.
Below we will determine the
integration constants $\Sigma^{(1)}$ and $\Sigma^{(2)}$ by comparison
with the dual field theory.  The presence of $\Delta_{\rho}$, on the other
hand, is crucial in order to obtain invariants under rescalings of $r$,
as we will discuss below.

By setting $\Delta_V = \Sigma^{(1)} - 2 \Lambda \, a_0
\Big(1- \sqrt{\frac{3 }{2}} \,
\frac{a_1}{a_0^2} \Big) $ we
note that (\ref{sol2}) reduces to (\ref{sol1}) to linear order
in $\Lambda$.

Inserting (\ref{sol2}) into (\ref{f2sec}) yields
\bea
f = \left(2 \Lambda  a_0  \Big(1- \sqrt{\frac{3 }{2}}
\frac{a_1}{a_0^2} \Big)
+ \frac{\Lambda \Sigma^{(1)}}{a_0}
\Big(1+ \sqrt{\frac{3 }{2}}
\frac{a_1}{a_0^2} \Big)  - 6 \frac{\Lambda
\Delta_{\rho}}{a_0}
-  4 \Lambda  a_0 \Big(1 +  \frac{\Sigma^{(1)}}{2 a_0^2} \Big)
r \right)
 {\rm e}^{-2r} .
\label{fc2}
\eea
Then we compute
\bea
f + \delta V = \left( a_0^2 + \Sigma^{(1)} + \Sigma^{(2)}
 \right) \, {\rm e}^{-2r} \;.
\eea
Using (\ref{simpl}) we obtain
\bea
\tau = a_0 \left(1 + \frac{\Sigma^{(1)}}{2 a_0^2} -
\frac{(\Sigma^{(1)})^2}{8 a_0^4} + \frac{\Sigma^{(2)}}{2 a_0^2}
\right)\, {\rm e}^{-r} \;.
\label{tau}
\eea
If we demand that there are no terms proportional to $\Lambda^2$
in (\ref{tau}), then we infer that
\bea
\Sigma^{(2)} = \frac{(\Sigma^{(1)})^2}{4 a_0^2}  \;.
\label{sigma12}
\eea
In the next section we will discuss the dual field theory realisation
yielding (\ref{sigma12}).

Using (\ref{sigma12}), we thus find
that to order $\Lambda^2$, and near the UV fixed point
$gW=1$, the curved domain wall solution is supported
by the three scalar fields
\bea
\rho &=& 1 + \left( \frac{2}{3} a_0^2 \,\Big(1 +
\frac{\Sigma^{(1)}}{2 a_0^2}
\Big)^2 \,r + \frac{a_1}{\sqrt{6}}
+ \Delta_{\rho} \right) \,
{\rm e}^{-2r} \;, \nonumber\\
f &=& \left(
-  4 \Lambda  a_0 \Big(1 +  \frac{\Sigma^{(1)}}{2 a_0^2} \Big)
r +
2 \Lambda  a_0  \Big(1- \sqrt{\frac{3 }{2}}
\frac{a_1}{a_0^2} \Big)
+ \frac{\Lambda \Sigma^{(1)}}{a_0}
\Big(1+ \sqrt{\frac{3 }{2}}
\frac{a_1}{a_0^2} \Big)  - 6 \frac{\Lambda
\Delta_{\rho}}{a_0}
 \right)
 {\rm e}^{-2r} \;,
\nonumber\\
\tau &=& a_0 \left(1 + \frac{\Sigma^{(1)}}{2 a_0^2}
\right)\, {\rm e}^{-r} \;.
\label{solul2}
\eea
We note that
in deriving
(\ref{solul2}) we worked to order ${\rm e}^{-2r}$.

The presence of the integration constant $a_1$ in the flat domain
wall solution (\ref{flats}) is necessary for the solution to be
invariant under additive shifts of $r$ ($r \rightarrow r + \beta$)
\cite{FGPW}.  Under such an additive shift, $a_0$ and $a_1$ transform
as $a_0 \rightarrow {\rm e}^{\beta} a_0, a_1 \rightarrow
 {\rm e}^{\beta} (a_1 - \sqrt{\ft83} a_0^2\, \beta)$.  The
invariant combination is thus given by $a_1/a_0^2
+ \sqrt{\ft83} \log a_0$,
and its value is determined by demanding that the RG flow terminates at a
superconformal fixed point in the IR \cite{FGPW}.  In the curved case,
on the other hand, we have $\Lambda \rightarrow {\rm e}^{\beta}
\Lambda, \Sigma^{(1)} \rightarrow {\rm e}^{2\beta} \Sigma^{(1)}$
under additive shifts of $r$.  Then, to order $\Lambda^2$,
the solution (\ref{solul2}) is
invariant under these additive shifts provided that
$\Delta_{\rho} \rightarrow {\rm e}^{2\beta} [ \Delta_{\rho}
- \ft23 (\Sigma^{(1)} + \ft14 (\Sigma^{(1)})^2/a_0^2)\, \beta ] $.

The
three non-constant scalar fields
$\rho, f$ and $\tau$ should be in
one-to-one correspondence with the deformations in the dual field theory.
$\delta \rho$ and $f$, which contain terms proportional to
$r \, {\rm e}^{-2r}$, should correspond
to bosonic mass deformations, whereas $\tau$, which behaves
as ${\rm e}^{-r}$, should correspond to a fermionic mass deformation
\cite{Witten:1998qj,KlWi}.  We will, in the next
section, show that this is indeed the case.
In doing so, we will determine the
integration constant $\Sigma^{(1)}$.

\section{The dual field theory on a curved background \label{sec:dual}}

\setcounter{equation}{0}

The dual field theory studied by Freedman, Gubser, Pilch and Warner
consists of
${\cal N}=4$ super-Yang-Mills theory broken to an ${\cal N}=1$
theory by the addition of a mass term for one of the three adjoint
chiral superfields.  This field theory
may be put on a curved background by setting
the gravitational
superfield ${\cal R}$ to a constant value and by setting
$W_{\alpha \beta \gamma} = G_{\alpha \dot{\alpha}} =0$
(see \cite{Gates} and references therein).  This implies that
i) the auxiliary field $b_a$ vanishes,
ii)
the
lowest component field $M$ of ${\cal R}$ acquires a constant value, i.e.
$ M = 6 \Lambda {\rm e }^{i \alpha}$ and iii)
the curvature scalar $R$ becomes constant, $ R =
\ft43 {\bar M} M = 48 \Lambda^2$ (we use the notation of \cite{WB}).

The superfield Lagrangian describing the dual field theory on a curved
background is, in ${\cal N}=1$ superspace notation \cite{WB},
given by\footnote{There is also a cubic term
${\rm Tr} ( [\Phi_1, \Phi_2] \Phi_3 )$
in the superpotential $W$, which we will omit in the following.}
\bea
{\cal L} = \int d^2 {\Theta} \, 2 \varepsilon \, \left[ \frac{1}{4 k}
{\rm Tr} \, W^{\alpha}
W_{\alpha} - \frac{1}{8} ({\bar {\cal D}}_{\dot \alpha}
{\bar {\cal D}}^{\dot \alpha} - 8 {\cal R} ) {\Phi^{\dagger}} {\rm e}^V \Phi
+ \frac{m}{2}  {\rm Tr}\Phi^2 + c_1 \, {\cal R} \, {\rm Tr}\Phi^2
\right] + {\rm h.c.}.
\label{lagback}
\eea
Note that we have allowed for the presence of an additional holomorphic
term of the form ${\cal R} \, {\rm Tr}\Phi^2$, with an unspecified
dimensionless
coefficient $c_1$!

The component expansion of (\ref{lagback}) gives rise to a fermionic mass
term
\bea
(m - \frac{c_1}{3} M ) \, \chi \chi \;.
\label{ferm}
\eea
It also
gives rise to the following
auxiliary field contributions,
\bea
{\cal L}_{aux}/e &=& \frac{1}{9} {\bar A} A \,
|M - 3 \frac{{\bar F}}{\bar A}|^2
- \frac{1}{2} (m - \frac{c_1}{3} M)   {\bar M}
A^2
- \frac{1}{2} (m - \frac{c_1}{3} {\bar M}) M
{\bar A}^2 \nonumber\\
&&+ (m - \frac{c_1}{3} M)  A F +
(m - \frac{c_1}{3} {\bar M})
{\bar A} {\bar F} \;.
\label{aux}
\eea
Therefore, the equation of motion for the auxiliary field $F$ reads
\bea
{\bar F} = - (m - \frac{c_1}{3} M) A + \frac{1}{3} M {\bar A} \;.
\label{f}
\eea
Reinserting (\ref{f}) into (\ref{aux}) yields
\bea
{\cal L}_{aux}/e = - \left(
|m - \frac{c_1}{3} M|^2
 {\bar
A} A + \frac{1 }{6} (m - \frac{c_1}{3} M)
{\bar M} A^2 + \frac{1 }{6}
(m - \frac{c_1}{3} {\bar M})
M {\bar A}^2 \right) \;.
\eea
The kinetic part
${\cal L}_{kin}$ of the Lagrangian (\ref{lagback}),
 on the other hand, contains a term \cite{WB}
\bea
{\cal L}_{kin}/e = \frac{1}{6} R\, {\bar A} A + \dots = \frac{2}{9} {\bar M} M
\, {\bar A} A + \dots \;.
\eea
Thus, the resulting effective scalar potential is given by
\bea
V_{\rm eff} = \left(
|m - \frac{c_1}{3} M|^2 - \frac{2}{9} {\bar M} M \right)
 {\bar
A} A + \frac{1 }{6} (m - \frac{c_1}{3} M)
{\bar M} A^2 + \frac{1 }{6}
(m - \frac{c_1}{3} {\bar M})
M {\bar A}^2 \;.
\eea
In the following, we will only consider the case when
$M = \bar M = 6 \Lambda$, i.e. $\alpha =0$.  Then we obtain
\bea
V_{\rm eff} = \left(
m^2 - 4 c_1\, m \,\Lambda + 4(c_1^2 -2) \Lambda^2 \right)
 {\bar
A} A + \Lambda (m - 2 c_1 \,\Lambda)
( A^2 + {\bar A}^2 ) \;.
\label{pot}
\eea
Let us consider what happens when we set $m=c_1=0$.  Then the Lagrangian
(\ref{lagback}) describes ${\cal N}=4$ super-Yang-Mills theory on an
$AdS_4$ background, and we do not expect to have an RG flow.
The associated dual curved gravitational solution has constant scalars
($g W=1$).  The curved BPS equation for the warp factor now reads
(c.f. (\ref{floweqs}))
$U^{\prime} = \gamma = 1 - 2 \Lambda^2 {\rm e}^{-2U} + {\cal O}(\Lambda^4)$,
which is solved by ${\rm e}^U = {\rm e}^r + \Lambda^2 \,{\rm e}^{-r}$.
This corresponds to a curved slicing of $AdS_5$ \cite{DeWolfe:2000cp}.
Thus, the effect of the term $- 8 \Lambda^2
{\bar
A} A$ in $V_{\rm eff}$
is to induce a modification of the warp factor $U$.

On the other hand, when turning on either $m$ and/or $c_1$, we do induce an
RG flow.  We therefore expect to have a
correspondence between the non-constant scalar fields
(\ref{solul2}) supporting the curved domain wall and
the fermionic and the bosonic
deformations given in (\ref{ferm}) and in
\bea
{\tilde V}_{\rm eff} = V_{\rm eff} + 8 \Lambda^2 {\bar
A} A =
\left(
m - 2 c_1\,\Lambda \right)^2
 {\bar
A} A + \Lambda (m - 2 c_1 \,\Lambda)
( A^2 + {\bar A}^2 ) \;,
\label{pot2}
\eea
respectively.
We will, in the following, make the identification $a_0 =m$ \cite{FGPW}.
Inspection
of (\ref{solul2}) then shows that $\delta
\rho$  and $\tau$ are in correspondence
\cite{FGPW}
with the operators ${\bar A} A$ and
$\chi \chi$, respectively,
whereas $f$ is in correspondence
with the operator $A^2$.  Note that the latter gets switched off when
turning off the cosmological constant $\Lambda$.

Let us have a closer look at
the correspondence between $\tau$ and
$\chi \chi$.
The identification of $\tau$ given in (\ref{solul2}) with
(\ref{ferm}) yields
\bea
\Sigma^{(1)} = - 4\, c_1 \,a_0\, \Lambda \;.
\eea
Next, let us consider the correspondence of $f$ with $A^2$.
The coefficient multiplying the term $r {\rm e}^{-2r}$ in $f$ is proportional
to $\Lambda (a_0 + \ft12 \Sigma^{(1)}/a_0) = \Lambda ( a_0 - 2 c_1 \Lambda)$.
This is precisely the coefficient of the $A^2$ term in (\ref{pot2})!

And finally, let us consider the correspondence of $\delta
\rho$ with ${\bar A} A$.  The coefficient
multiplying the term $r {\rm e}^{-2r}$ in $\delta \rho$ is proportional
to
$(a_0 + \ft12
\Sigma^{(1)}/a_0)^2 =
(a_0 - 2\, c_1 \, \Lambda)^2$. This too is in precise
agreement with
 the coefficient of the ${\bar A} A$ term in
(\ref{pot2})!

We may add further perturbations to (\ref{lagback}), for instance
couplings of the form
\bea
\sum_{p \geq 2}^{\infty} c_p \,  {m^{1-p}} \,
\int d^2 {\Theta} \, 2 {\cal \varepsilon} \,
  {\cal R}^p \, {\rm Tr}\Phi^2
 + {\rm h.c.} \;
\eea
with dimensionless
coefficients $c_p$.  We expect that these can also be captured
by the curved domain wall solution.  Consider, for instance, adding
${\cal R}^2 \, {\rm Tr}\Phi^2$ ($p=2$).  This will result in additional
$\Lambda^2$ corrections to $\chi \chi$ and to ${\bar A} A$,
which can be captured on the domain wall side by replacing (\ref{sigma12})
by
\bea
\Sigma^{(2)} = \frac{(\Sigma^{(1)})^2}{4 a_0^2} + 4 c_2 \, \Lambda^2
=  4 (c_1 + c_2) \, \Lambda^2
\;.
\eea
To summarise, we have found that the dictionary between the scalar fields
supporting the curved domain wall solution and the deformations in the
dual field theory on an $AdS_4$ background is given by
\bea
\delta \rho \;\; &\longleftrightarrow& \;\; {\bar A} A \;, \nonumber\\
f  \;\; &\longleftrightarrow& \;\; A^2 \;, \nonumber\\
\tau \;\; &\longleftrightarrow& \;\; \chi \chi \;.
\eea

\section{Conclusions}
\setcounter{equation}{0}

In this note we put the
field theory studied by  Freedman, Gubser, Pilch and Warner on a curved
$AdS_4$ background, and we constructed the curved supersymmetric domain wall
solution which describes this field theory near the UV fixed point.
In doing so we allowed for the presence of additional deformations
in the field theory, for
instance of ${\cal R} {\rm Tr} \Phi^2$.
This example of a curved BPS domain wall
demonstrates that holographic RG flows in
supersymmetric field theories
on a curved $AdS_4$ background can be described in terms of curved BPS
domain wall solutions.

We constructed the curved domain wall solution to order $\Lambda^2$.
Inspection of (\ref{gamma}) shows that the
curved domain
wall solution will receive further corrections
which are higher order
in $\Lambda$ (i.e. of order $\Lambda^3, \Lambda^4$, etc.).
These will become important when flowing towards the infrared.
It would be interesting to investigate this further.

\bigskip

{\bf Acknowledgements}

We would like to thank G. Dall'Agata, J. Erdmenger, M. Faux, Z. Guralnik
and K. Landsteiner
for valuable discussions.

%%%%%%%%%%%%%%%%%%%%%%%%%%%%%%%%%%%%%%%%%%%%%%%%%%%%%%%%%%%%%%

% ---- Bibliography ----

%\nocite{*}                   %this uses *everything* in the .bib file
%\bibliography{8dDW}          %or whatever your .bib file is
%\bibliographystyle{utphys}   %if you use utphys.bst

%%%%%%%%%%%%%%%%%%%%%%%%%%%%%%%%%%%%%%%%%%%%%%%%%%%%%%%

\providecommand{\href}[2]{#2}\begingroup\raggedright

\endgroup

\end{document}